\documentclass[%
 reprint,
superscriptaddress,
 amsmath,amssymb,
 aps,
prb,
]{revtex4-1}

\usepackage{graphicx}
\usepackage{dcolumn}
\usepackage{bm}
\graphicspath{{Figures/}} 
\usepackage[caption=false]{subfig}
\usepackage{float}
\usepackage{multirow}
\usepackage[usenames]{color} 

\begin{document}

%

\preprint{APS/123-QED}

\title{Quantitative Measure of Hysteresis for Memristors Through Explicit Dynamics}

\author{P.S.~Georgiou}\email{ps.georgiou@imperial.ac.uk}
 \affiliation{Department of Chemistry,  Imperial College London}
\author{S.N. Yaliraki}
 \affiliation{Department of Chemistry,  Imperial College London}
\author{E.M. Drakakis}
 \affiliation{Department of Bioengineering,  Imperial College London}
\author{M. Barahona}\email{m.barahona@imperial.ac.uk}
 \affiliation{Department of Mathematics,  Imperial College London, South Kensington Campus, London SW7 2AZ, United Kingdom}

\date{\today}
\begin{abstract}
We introduce a mathematical framework for the analysis of the input-output dynamics of externally driven memristors. We show that, under general assumptions, their dynamics comply with a Bernoulli differential equation and hence can be nonlinearly transformed into a formally solvable linear equation. The Bernoulli formalism, which applies to both charge- and flux-controlled memristors when either current- or voltage-driven, can, in some cases, lead to expressions of the output of the device as an explicit function of the input. We apply our framework to obtain analytical solutions of the $i-v$ characteristics of the recently proposed model of the Hewlett-Packard memristor under three different drives without the need for numerical simulations.  Our explicit solutions allow us to identify a dimensionless lumped parameter that combines device-specific parameters with properties of the input drive. This parameter governs the memristive behavior of the device and, consequently, the amount of hysteresis in the $i-v$. We proceed further by defining formally a quantitative measure for the hysteresis of the device for which we obtain explicit formulas in terms of the aforementioned parameter and we discuss the applicability of the analysis for the design and analysis of memristor devices. 
\end{abstract}

\maketitle

\section{Introduction}
According to classical electrical circuit theory, there are three fundamental passive circuit elements: the resistor, the inductor and the capacitor. Back in 1971, Chua challenged this established perception~\cite{Chua1971}. He realized that only five out of the six possible pairwise relations between the four circuit variables (current, voltage, charge and magnetic flux) had been identified. Based on a symmetry argument, he postulated mathematically the existence of a fourth basic passive circuit element that would establish the missing link between charge and magnetic flux. The postulated element was named the \textit{memristor} because, unlike a conventional ohmic element, its instantaneous resistance depends on the entire history of the input (voltage or current) applied to the component~\cite{Oster1972,Oster1974}. Hence the memristor can be understood in simple terms as a non-linear resistor with memory. Chua and Kang~\cite{Chua1976,Chua1980} generalized this concept to a broader family of non-linear dynamical systems, which they termed \textit{memristive}, and demonstrated their usefulness in the modeling and understanding of physical and biological systems such as discharge tubes, the thermistor, or the Hodgkin-Huxley circuit model of the neuron. Although in the intervening years experimental devices with characteristics similar to the memristor (i.e., hysteresis, zero crossing and memory) were investigated, researchers generally failed to associate or explain them in the context of memristive systems~\cite{Baatar2009}.

The memristor remained an elusive theoretical component until recently, when scientists at Hewlett-Packard (HP) fabricated a solid-state device which was recognized as a memristor~\cite{Williams2008,Strukov2008}. In that work, Williams, Strukov and co-workers also provided a simple heuristic model to understand how memristive behavior could emerge from the underlying physical process. The report of the fabrication of the HP memristor reignited the interest of the community in such circuits, due to their potential widespread applications. This has resulted in various attempts to build memristor devices based on different underlying physical principles, as well as efforts to understand pre-existing devices in the context of memristive systems. Examples include thin-films \cite{Borghetti2009a,Pickett2009,Strukov2009b,Williams2008,Yang2008,Driscoll2009a}, nano-particle assemblies~\cite{Kim2009}, spintronics \cite{Pershin2008} or neurobiological systems \cite{Pershin2009,Baranco2010}. On the theoretical front, however, the analysis of these systems has been mostly restricted to numerics (i.e., temporal integration of model differential equations and numerical sweeping of parameters) due to the absence of a general mathematical framework that can provide analytical solutions to their dynamics.  

Here, we introduce a mathematical framework for the study of memristors that allows us to provide analytic solutions for their dynamics and input-output characteristics.  The framework is based on the compliance of a general class of ideal memristor dynamics with Jacob Bernoulli's differential equation, a classic nonlinear equation that can be solved analytically.  This formulation provides a powerful and systematic methodology for the analysis, characterization and design of devices governed by Bernoulli dynamics that does not rely on computationally expensive sweeping of parameters~\cite{Manos_thesis,Drakakis2000,Drakakis2010}. Moreover, our analytical results can be used to reveal the relevant combinations of parameters that govern the response characteristics of the memristor. 

The paper is structured as follows. Firstly, we introduce the general theoretical framework for memristor dynamics and the analytical solutions that follow from the Bernoulli differential equation. We then illustrate its application through the HP memristor~\cite{Strukov2008}, which is shown to comply with the Bernoulli framework, and we obtain analytical expressions of the output of the model when excited by three fundamental inputs: sinusoidal, bipolar square and triangular waves. To exploit further the insight provided by the framework, we define a quantitative measure of the hysteresis of the memristor in terms of the work done by the driving signal, and we use the derived analytical solutions to show that the hysteresis of the device depends on a specific dimensionless lumped quantity that combines all the parameters of the model.  This lumped parameter relates fabrication and device properties together with characteristics of the input signal and shows how such experimental parameters can be used to design and control the response of the memristor. 

\section{Memristor Dynamics under time-varying inputs}

\subsection{Definitions and background} \label{Definitions}

Consider the four fundamental circuit variables: voltage $v$, current $i$, charge $q$, and magnetic flux $\varphi$. There are six distinct relations linking these variables pairwise. Two of these relations correspond to the definitions of charge and magnetic flux as time-integrated variables:
\begin{eqnarray*}
q(t)= \int_{-\infty}^t i(\tau)d\tau\\ 
\varphi(t)=\int_{-\infty}^t v(\tau) d\tau.
\end{eqnarray*} 
Three other links are given by the implicit equations that define the constitutive laws of the generalized fundamental circuit elements:
\begin{eqnarray}
f_\mathcal{R}(v,i)&=&0 \quad \text{for the resistor,}   \label{eq:implicit_resistor}\\ 
f_\mathcal{C}(v,q)&=&0 \quad \text{for the capacitor,}\\ 
f_\mathcal{L}(\varphi,i)&=&0 \quad \text{for the inductor.}
\end{eqnarray}
In order to complete the symmetry of the system-theoretic structure, Chua's insight was to postulate that the remaining link between $q$ and $\varphi$ should be completed by another constitutive relation 
\begin{equation} 
f_\mathcal{M}(\varphi,q)=0,
\label{eq:implicit_memristor}
\end{equation}
which would correspond to a missing element: the \emph{memristor}. In this sense, the memristor complements the other three fundamental circuit elements as the fourth ideal passive two-terminal component~\cite{Chua1980,Chua1971,Oster1972}.

Assume a time-invariant memristor with no explicit dependence on time, i.e., $\partial{f_\mathcal{M}}/\partial{t} = 0$. Then the time derivative of the constitutive relation~(\ref{eq:implicit_memristor}) leads to
\begin{equation}
 \label{eqn_sec2_dfm_dt}
 \frac{d f_\mathcal{M}}{dt} = \left ( \frac{\partial f_\mathcal{M}}{\partial \varphi} \right )_{q}  \frac{d\varphi}{dt} + \left ( \frac{\partial f_\mathcal{M}}{\partial q} \right )_{\varphi} \frac{dq}{dt}= 0,
\end{equation}
which serves us to define the memristance $\mathcal{M}(q,\varphi)$~\cite{Chua1971,Oster1972}:
\begin{equation}
 \label{eqn_sec2_memr_def_i-v}
 v = - \frac{\left (\frac{\partial f_\mathcal{M}}{\partial q}\right )_{\varphi}} {\left (\frac{\partial f_\mathcal{M}}{\partial \varphi}\right)_{q}} \, i(t) \equiv \mathcal{M}(q,\varphi)\, i.
\end{equation}
Since we are considering \emph{strictly passive} elements, with $\mathcal{M}(q,\varphi) > 0$, the constitutive law~(\ref{eq:implicit_memristor}) defines a strictly monotonically increasing function in the $(q,\varphi)$ plane. It then follows that we can express the memristor constitutive law as an explicit function either in terms of the flux, $q=q_\mathcal{M}(\varphi)$, or in terms of the charge, $\varphi=\varphi_\mathcal{M}(q)$, indistinctly and to our convenience~\cite{Apostol1957}. 

Take, for instance, a current-driven memristor, in which $i(t)$ is the time-dependent input under our control, then we can formally express the input-output relationship in terms of $q$, an integrated variable of the input $i(t)$ over its past:
\begin{equation}
 v = \mathcal{M}\left(\int_{-\infty}^{t} i(\tau) d \tau \right ) \, i(t), \label{eq:memristor_memory}
\end{equation}
a form that makes explicit the memory of the device.

Note, in contrast, that a time-invariant resistor governed by~(\ref{eq:implicit_resistor}) has a differential resistance $\mathcal{R}(v,i)$ 
\begin{equation}
\label{eqn_sec2_res_def}
dv = - \frac{\left (\frac{\partial f_\mathcal{R}}{\partial i}\right )_{v}} {\left (\frac{\partial f_\mathcal{R}}{\partial v}\right)_{i}} di \equiv \mathcal{R}(v,i) \, di.
\end{equation} 
The input-output relationship of a current-driven passive resistor is then given by
\begin{equation}
 v = \int_{-\infty}^{i} \mathcal{R}(\eta) d\eta,  \label{eq:resistor_nomemory}
\end{equation}
which is in general nonlinear but, unlike Eq.~(\ref{eq:memristor_memory}), it exhibits no memory.

Hence, the memory property follows from the integration necessary to go from the $i-v$ plane, which is readily accessible to experiments, to the $q-\varphi$ plane in which the memristor constitutive law is defined~\cite{Chua1971,Oster1974}. The memristance~(\ref{eqn_sec2_memr_def_i-v}) and the resistance~(\ref{eqn_sec2_res_def}) are thus dimensionally congruent but they only coincide if they are both constant, i.e., for the ohmic (linear and constant) resistor~\cite{Chua1971}.

\subsection{Solution of memristor dynamics under time-dependent inputs: the Bernoulli equation} \label{General Solution}

Although the constitutive relation of the memristor is defined in the $q-\varphi$ plane, such devices are characterized experimentally in the $i-v$ plane. Specifically, it is typical to characterize their output dynamics  in response to time-dependent inputs. The particular relationship imposed by the definition of the memristor translates into a restricted form of the differential equations that govern the dynamics in the $i-v$ time domain. In particular, the dynamics obey a Bernoulli differential equation (BDE) that can be reduced to an associated linear time-dependent differential equation (LDE). In some cases, this structure can be exploited to provide analytical solutions for the input-output dynamics without the need for explicit numerical integration of the model. 

The Bernoulli equation has the form~\cite{Handbook_of_ODEs}:
\begin{equation}
\label{eqn_sec2_bernoulli_general}
\frac{dy}{dx} + f(x)y=g(x)y^n 
\end{equation}
where $n \in \mathbb{N}$. If $n=\{0, 1\}$, the equation is linear. For $n>1$ this family of nonlinear ODEs can be solved analytically using the nonlinear change of variable  $z=y^{1-n}$ to give the LDE
\begin{equation*}
 \frac{dz}{dx} + (1-n)f(x)z=(1-n)g(x)
\end{equation*}
from which the general solution follows:
\begin{equation}
 \label{eqn_sec2_bernoulli_general_solution}
  y =\left [ \frac{(1-n) \, \int m(x)g(x)dx + C}{m(x)} \right ]^{\frac{1}{1-n}}.
\end{equation}
Here $m(x)= e^{(1-n)\int f(x) dx}$ is an integrating factor and $C$ is the constant of integration.

In the case of memristor dynamics, the mathematical transformation that relates the BDE/LDE pair is not only a convenient manipulation but is also associated with the experimental measurement setup and the intrinsic description of the  memristor: both the BDE and LDE can appear depending on the choice of externally-driven variable ($i$ or $v$) in conjunction with the internal variable ($q$ or $\varphi$) through which the memristive behavior is controlled.

In particular, consider an experiment in which the memristor is described in terms of the integrated input, e.g., a current-driven memristor with input $i(t)$ whose memristance is charge-controlled, $\mathcal{M}(q)$. The dynamical system can then be written in the canonical form of a memristive system:
\begin{eqnarray}
 \dot{q} &=& i(t)  \nonumber \\ 
 v &=& \mathcal{M}(q) \, i(t) \label{eqn_sec2_memr_def_part2}.
\end{eqnarray}
This leads directly to the following LDE:
\begin{equation}
 \label{eqn_sec2_linear_}
 \frac{dv}{dt} - \left [ \frac{1}{i(t)}\frac{di(t)}{dt} \right ] {v}=  \left [ \frac{d \mathcal{M}}{dq} \right ] i(t)^2,
\end{equation}
which has the standard analytical solution shown in Table~\ref{tbl_ch3_general_solutions}.

However, in certain situations, it might be more meaningful (or convenient) to express the memristance as a function of the integrated \textit{dependent} variable. For instance, if the memristor is voltage-driven with input $v(t)$ but the memristance is given in terms of the charge $\mathcal{M}(q)$, we have
\begin{eqnarray}
 \dot{\varphi} &=& v(t)  \label{eqn_sec2_memr_voltage1} \nonumber \\
 \mathcal{M}(q) \, i &=& v(t) \label{eqn_sec2_memr_voltage2}
 \label{eq:Bernoulli_dependent}
\end{eqnarray}
and the resulting equation is:
\begin{equation}
 \label{eqn_sec2_bernoulli_2}
 \frac{di}{dt} - \left [ \frac{1}{v(t)}\frac{dv(t)}{dt} \right ] {i}= - \left [  \frac{1}{v(t)} \frac{d\mathcal{M}}{dq} \right ] i^3,
\end{equation}
which has a Bernoulli form. Indeed, this is the case for the first HP memristor model~\cite{Strukov2008}, as shown in the next section.

In general, the output dynamics of the general memristor leads to four cases. Differentiating \eqref{eqn_sec2_memr_def_i-v} with respect to time yields
\begin{equation}
\frac{dv}{dt} = \mathcal{M}\frac{di}{dt}+ i \left [ \frac{\partial{\mathcal{M}}}{\partial{q}} i + \frac{\partial{\mathcal{M}}}{\partial{\varphi}} v \right ].
\label{eq:memristor_dif_t}
\end{equation}
If the device is current-driven, the dynamical equations are:  
\begin{subequations}
 \label{eqn_ch3_iMem_dyn}
\begin{align}
 \frac{dv}{dt}-\left[\frac{1}{i(t)}\frac{di(t)}{dt}\right]v &= \left[\frac{\partial\mathcal{M}}{\partial q} + \frac{\partial\mathcal{M}}{\partial \varphi}\mathcal{M}\right] i(t)^2 \\
 &= \left[\frac{\partial\mathcal{M}}{\partial q}\frac{1}{\mathcal{M}^2} + \frac{\partial\mathcal{M}}{\partial \varphi}\frac{1}{\mathcal{M}} \right] v^2,
\end{align}
\end{subequations}
while for the voltage-driven memristor we have:
\begin{subequations}
 \label{eqn_ch3_vMem_dyn}
 \begin{align}
 \frac{di}{dt} - \left[\frac{1}{v(t)}\frac{dv(t)}{dt}\right] i &= -\left[\frac{\partial\mathcal{M}}{\partial q}\frac{1}{\mathcal{M}^3} + \frac{\partial\mathcal{M}}{\partial \varphi}\frac{1}{\mathcal{M}^2} \right]v(t)^2\\
 &=-\left[\frac{\partial\mathcal{M}}{\partial q}\frac{1}{\mathcal{M}} + \frac{\partial\mathcal{M}}{\partial \varphi} \right] i^2,
 \end{align}
\end{subequations}
where we have emphasized in the equations the explicit dependence on time of the externally controlled drives. Note that the dynamics are formally either a  BDE (Eqs.~(\ref{eqn_ch3_iMem_dyn}b) and~(\ref{eqn_ch3_vMem_dyn}b)) or a LDE (Eqs.~(\ref{eqn_ch3_iMem_dyn}a) and~(\ref{eqn_ch3_vMem_dyn}a)), which are related through the nonlinear transformation given above.  For \emph{strictly passive} memristors~\cite{Chua1971,Chua1976} (for which $\mathcal{M}(q)$, or equivalently $\mathcal{M}(\varphi)$, is always positive), the memristance can be represented as a function of either the independent or dependent variable, as convenient, due to the existence of the inverse function~\cite{Apostol1957}. Therefore, once we specify the intrinsic integrated variable (i.e., $q$ or $\varphi$) that controls the specific memristor, we can obtain BDEs and LDEs as summarized in Table~\ref{tbl_ch3_general_solutions} with their corresponding general solutions. 

\begin{table*}[htb]
\renewcommand{\arraystretch}{1.3}
\centering
\begin{tabular}{cccll}

\multicolumn{2}{c}{\textbf{Memristor Type}} & \multicolumn{2}{c}{\textbf{Governing Differential Equation}} & \multicolumn{1}{c}{\textbf{General Solution}} \\
\hline\hline
\multirow{8}{1.6cm}{\centering Charge Controlled} & \multirow{4}{1.6cm}{\centering Current Driven} & \multirow{2}{1cm}{\centering LDE} & \multirow{2}{*}{$\dfrac{dv}{dt}-\left[\dfrac{1}{i}\dfrac{di}{dt}\right]v=\left[\dfrac{d\mathcal{M}(q)}{dq}\right]i^2 $}                                                  & \multirow{2}{*}{$v(t)=i(t)\left[\mathcal{M}_0 + \displaystyle\int_0^t \dfrac{d\mathcal{M}(q)}{dq}i(\tau)d\tau\right] $} \\
                                                  &                                                &                                   &                                                                                                                                                                          & \\
                                                  &                                                & \multirow{2}{1cm}{\centering BDE} & \multirow{2}{*}{$\dfrac{dv}{dt}-\left[\dfrac{1}{i}\dfrac{di}{dt}\right]v=\left[\dfrac{d\mathcal{M}(q)}{dq}\dfrac{1}{\mathcal{M}^2(q)}\right]v^2 $}                       & \multirow{2}{*}{$v(t)=i(t)\left[\mathcal{M}_0^{-1}-\displaystyle\int_0^t \dfrac{d\mathcal{M}(q)}{dq} \dfrac{i(\tau)}{\mathcal{M}^2(q)}d\tau\right]^{-1}$}\\
                                                  &                                                &                                   &                                                                                                                                                                          & \\
\cline{3-5}
                                                  & \multirow{4}{1.6cm}{\centering Voltage Driven} & \multirow{2}{1cm}{\centering LDE} & \multirow{2}{*}{$\dfrac{di}{dt}-\left[\dfrac{1}{v}\dfrac{dv}{dt}\right]i=-\left[\dfrac{d\mathcal{M}(q)}{dq}\dfrac{1}{\mathcal{M}^3(q)}\right]v^2 $}                      & \multirow{2}{*}{$i(t)=v(t)\left[\mathcal{M}_0^{-1}-\displaystyle\int_0^t\dfrac{d\mathcal{M}(q)}{dq}\frac{v(\tau)}{\mathcal{M}^3(q)}d\tau\right] $}\\
                                                  &                                                &                                   &                                                                                                                                                                          & \\
                                                  &                                                & \multirow{2}{1cm}{\centering BDE} & \multirow{2}{*}{$\dfrac{di}{dt}-\left[\dfrac{1}{v}\dfrac{dv}{dt}\right]i=-\left[\dfrac{d\mathcal{M}(q)}{dq}\dfrac{1}{v}\right]i^3 $}                                     & \multirow{2}{*}{$i(t)=v(t)\left[\mathcal{M}_0^2+2\displaystyle\int_0^t\dfrac{d\mathcal{M}(q)}{dq}v(\tau)d\tau\right]^{-\frac{1}{2}} $}\\
                                                  &                                                &                                   &                                                                                                                                                                          & \\

\cline{2-5}
 \multirow{8}{1.6cm}{\centering Flux Controlled}  & \multirow{4}{1.6cm}{\centering Current Driven} & \multirow{2}{1cm}{\centering LDE} & \multirow{2}{*}{$\dfrac{dv}{dt}-\left[\dfrac{1}{i}\dfrac{di}{dt}\right]v=\left[\dfrac{d\mathcal{M}(\varphi)}{d\varphi}\mathcal{M}(\varphi)\right]i^2 $}                  & \multirow{2}{*}{$v(t)=i(t)\left[\mathcal{M}_0+\displaystyle\int_0^t\dfrac{d\mathcal{M}(\varphi)}{d\varphi}\mathcal{M}(\varphi)i(\tau)d\tau\right] $}\\
                                                  &                                                &                                   &                                                                                                                                                                          & \\
                                                  &                                                & \multirow{2}{1cm}{\centering BDE} & \multirow{2}{*}{$\dfrac{dv}{dt}-\left[\dfrac{1}{i}\dfrac{di}{dt}\right]v=\left[\dfrac{d\mathcal{M}(\varphi)}{d\varphi}\dfrac{1}{\mathcal{M}(\varphi)}\right]v^2 $}       & \multirow{2}{*}{$v(t)=i(t)\left[\mathcal{M}_0^{-1}-\displaystyle\int_0^t\dfrac{d\mathcal{M}(\varphi)}{d\varphi}\frac{i(\tau)}{\mathcal{M}(\varphi)}d\tau\right]^{-1} $}\\
                                                  &                                                &                                   &                                                                                                                                                                          & \\
\cline{3-5}
                                                  & \multirow{4}{1.6cm}{\centering Voltage Driven} & \multirow{2}{1cm}{\centering LDE} & \multirow{2}{*}{$\dfrac{di}{dt}-\left[\dfrac{1}{v}\dfrac{dv}{dt}\right]i=-\left[\dfrac{d\mathcal{M}(\varphi)}{d\varphi}\dfrac{1}{\mathcal{M}^2(\varphi)}\right]v^2 $}    & \multirow{2}{*}{$i(t)=v(t)\left[\mathcal{M}_0^{-1}-\displaystyle\int_0^t\dfrac{d\mathcal{M}(\varphi)}{d\varphi}\frac{v(\tau)}{\mathcal{M}^2(\varphi)}d\tau\right] $}\\
                                                  &                                                &                                   &                                                                                                                                                                          & \\
                                                  &                                                & \multirow{2}{1cm}{\centering BDE} & \multirow{2}{*}{$\dfrac{di}{dt}-\left[\dfrac{1}{v}\dfrac{dv}{dt}\right]i=-\left[\dfrac{d\mathcal{M}(\varphi)}{d\varphi}\right]i^2 $}                                     & \multirow{2}{*}{$i(t)=v(t)\left[\mathcal{M}_0+\displaystyle\int_0^t\dfrac{d\mathcal{M}(\varphi)}{d\varphi}v(\tau)d\tau\right]^{-1}$}\\
                                                  &                                                &                                   &                                                                                                                                                                          & \\                                           
\hline
\end{tabular}
\caption[General Solutions]{Governing differential equations and corresponding general solutions for different types of memristors, as characterized by their controlling variable, when driven by different inputs.}
\label{tbl_ch3_general_solutions}
\end{table*}

The above discussion shows that the constitutive relationship of the memristor leads generally to Bernoulli dynamics which can be reduced to the corresponding time-varying linear dynamics through a nonlinear transformation. The analytical solutions obtained can be used to understand explicitly the parameter dependence of memristor models as well as to aid in the design process without the need for parameter sweeping via numerical simulations of the dynamics. In particular, we obtain below analytical expressions for the behavior of a memristor model when driven with different input waveforms.  Furthermore, our analytical solutions allow us to show that the behavior of the system is fully characterized by a particular lumped parameter that incorporates all the physical parameters of the memristor. This renormalized parameter is shown to encapsulate both the hysteretic and nonlinear characteristics of the device and its memristive character.  

\subsection{Dynamics of the HP memristor under time-varying inputs}

We exemplify our approach with the HP model, which was introduced by Strukov \textit{et al.} to describe the behavior of their fabricated memristor~\cite{Strukov2008}. Their device consists of a thin-film semiconductor of $\text{TiO}_2$ with thickness $D$ placed between two metal contacts made of $\text{Pt}$. The film has two regions: a doped region of thickness $w$ with low resistivity $R_{ON}$ due to the high concentration of dopants (positive oxygen ions) and an undoped region with thickness $(D-w)$ and high resistivity $R_{OFF}$. The total resistance of the film is modeled as two variable resistors in series whose resistance depends on the position of the boundary between the doped and undoped regions. Applying an external voltage across the device will have the effect of moving the boundary between the doped and undoped region effectively changing the total resistance of the device. Assuming ohmic electronic conductance and linear ionic drift with average vacancy mobility $\mu_v$,  the device is modeled by the following pair of equations~\cite{Strukov2008}:
\begin{eqnarray}
 \label{eqn_sec2_williams1_v}
 v(t) &=& \left[ \mathcal{R}_{ON} \frac{w(t)}{D} + \mathcal{R}_{OFF}\left(1 - \frac{w(t)}{D}\right)\right] i(t) \\
 \label{eqn_sec2_Williams1_w}
 \dot{w}(t) &=& \mu_v \frac{\mathcal{R}_{ON}}{D} \, i(t).
\end{eqnarray}
Equation~(\ref{eqn_sec2_Williams1_w}) means that the position of the boundary between regions is proportional to the total charge that passes through the device. Hence Eq.~(\ref{eqn_sec2_williams1_v})  is a memristor relationship of the form given in Eq.~(\ref{eq:Bernoulli_dependent}) with memristance controlled by charge~\footnote{Note that  $k_2$, the corresponding constant in Ref.~\cite{Georgiou2011},  is the negative of $\kappa$ here: $\kappa=-k_2$.}:
\begin{eqnarray}
 \mathcal{M}(q)&=& \mathcal{R}_{OFF}+ \mu_v \left(\frac{\mathcal{R}_{ON}}{D}\right)^2  \left(1-\frac{\mathcal{R}_{OFF}}{\mathcal{R}_{ON}}\right) q \nonumber \\ 
 &\equiv&  \mathcal{R}_{OFF} - \kappa \, q .
 \label{eqn_sec2_HP_memristance}
 \end{eqnarray}
When the device is voltage-driven (as is the case in the original model~\cite{Strukov2008}) the dynamics is governed by Eq.~\eqref{eqn_sec2_bernoulli_2}. A simple inspection of Table~\ref{tbl_ch3_general_solutions} shows that the solution of the corresponding BDE leads to an $i-v$ relationship with the following explicit form:
\begin{equation}
 \label{eqn_sec2_Williams1_solution}
 i = \frac{v(t)}{\sqrt{R_0^2 - 2 \kappa \int_0^t v(\tau)d\tau}},
\end{equation}
where the constant of integration $  R_0 \in [\mathcal{R}_{ON},\mathcal{R}_{OFF}] $ is the resistance of the device at $t_0$, the initial time at which the input signal is applied.

\begin{figure*}[t]
 \begin{center}
  \subfloat[]{\label{fig_Ix_Vx_R0_a}\includegraphics[width=0.33\textwidth]{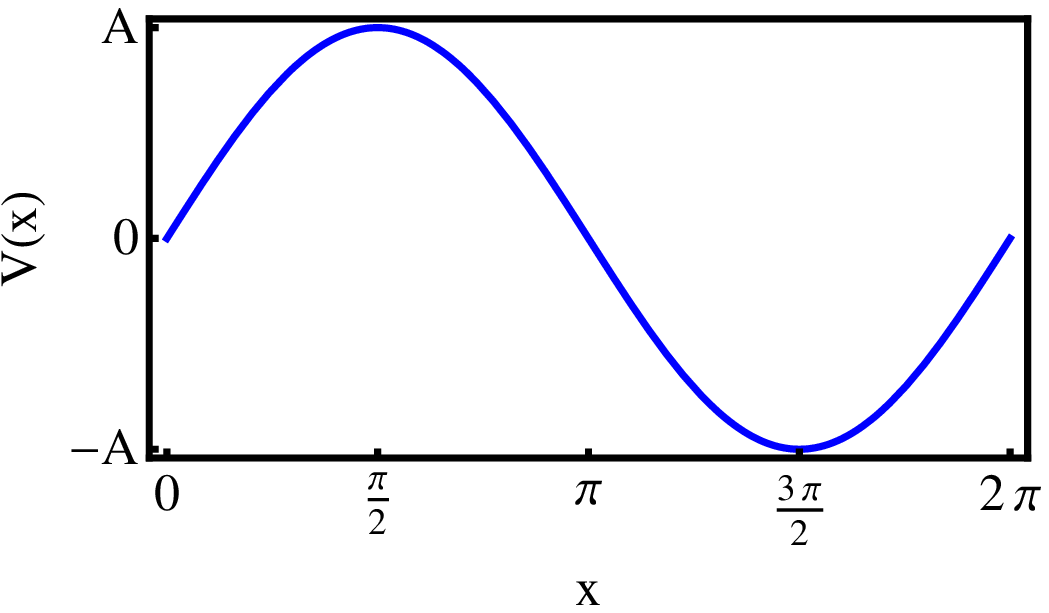}}                
  \subfloat[]{\label{fig_Ix_Vx_R0_b}\includegraphics[width=0.33\textwidth]{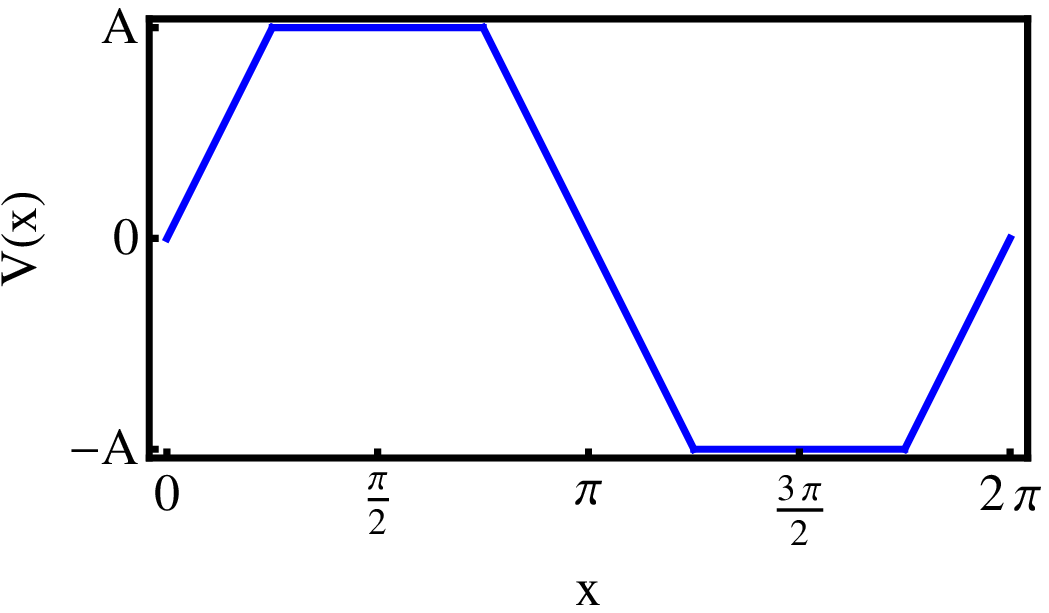}}
  \subfloat[]{\label{fig_Ix_Vx_R0_c}\includegraphics[width=0.33\textwidth]{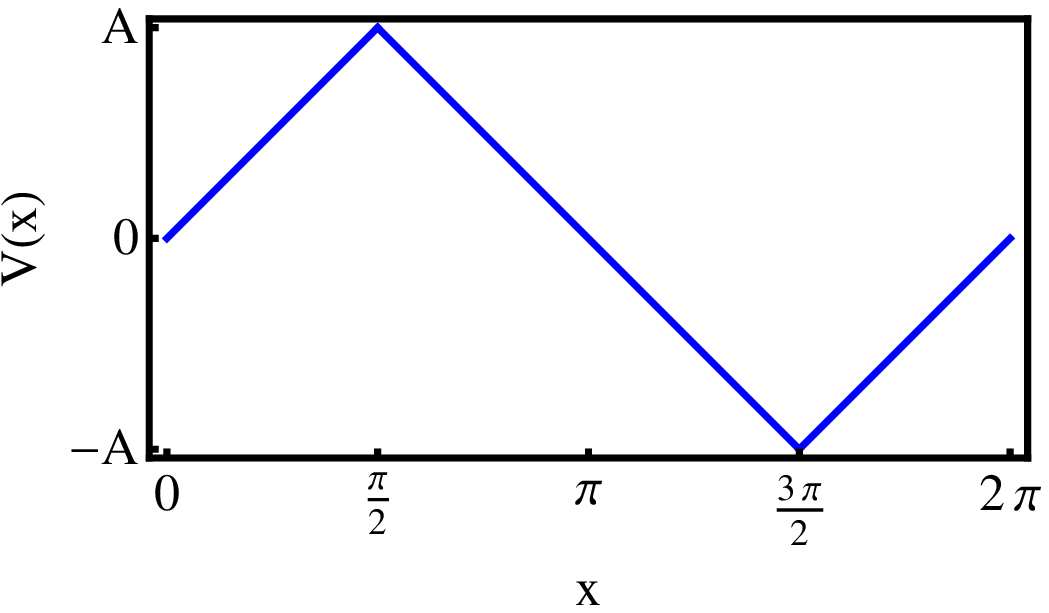}}\\
  \subfloat[]{\label{fig_Ix_Vx_a}\includegraphics[width=0.33\textwidth]{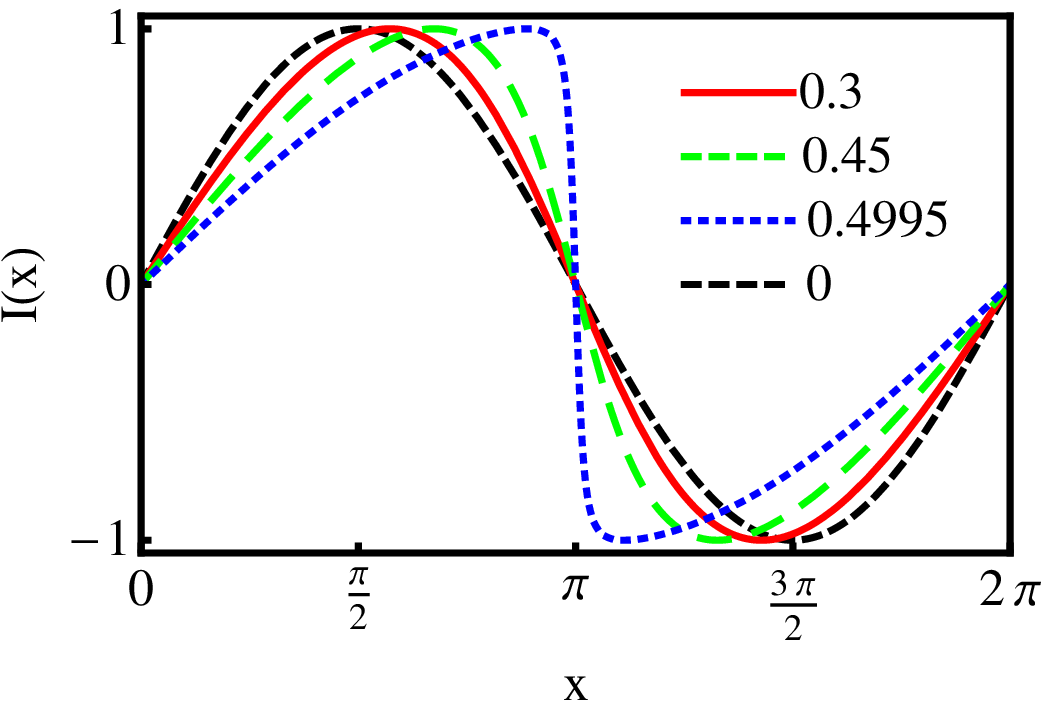}}                
  \subfloat[]{\label{fig_Ix_Vx_b}\includegraphics[width=0.33\textwidth]{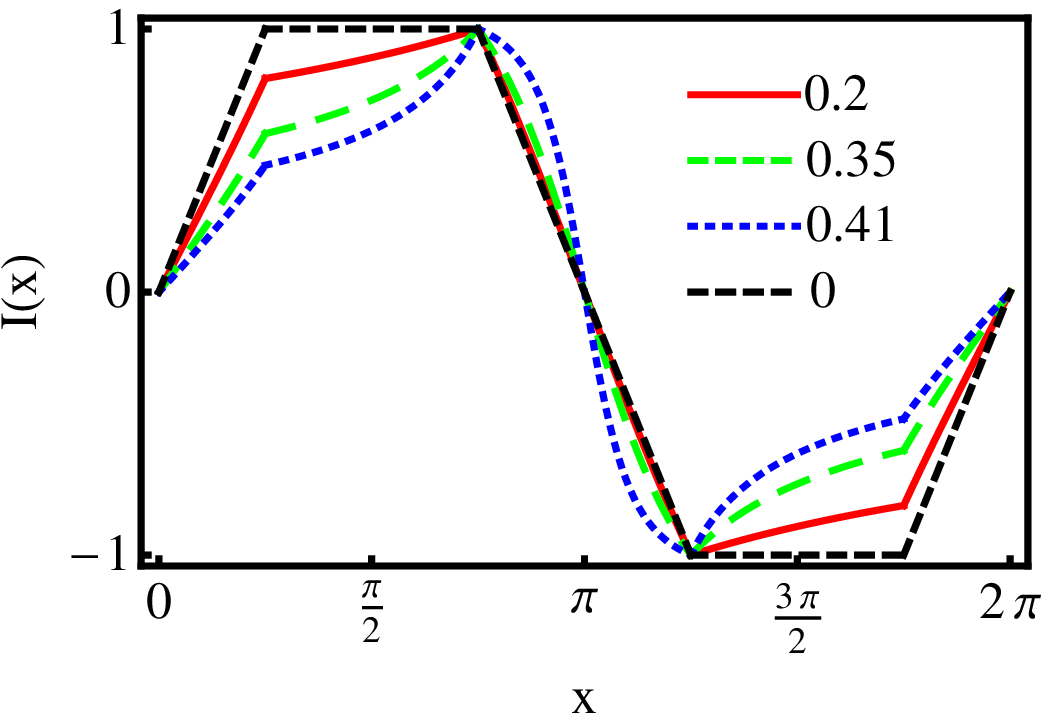}}
  \subfloat[]{\label{fig_Ix_Vx_c}\includegraphics[width=0.33\textwidth]{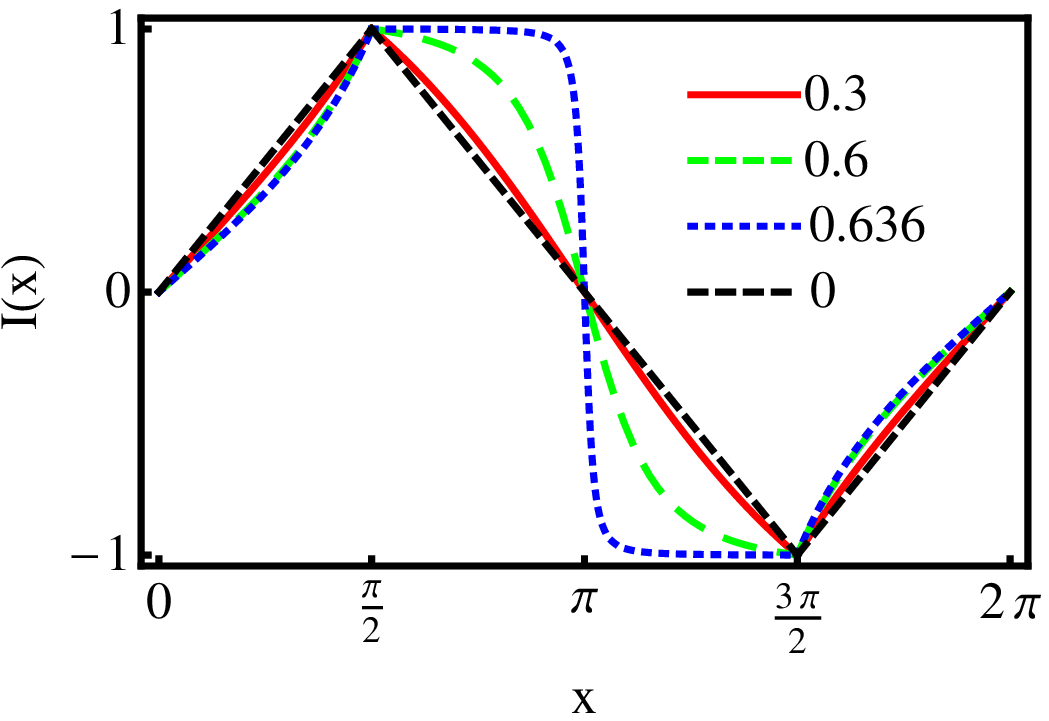}}\\
  \subfloat[]{\label{fig_I_V_a}\includegraphics[width=0.33\textwidth]{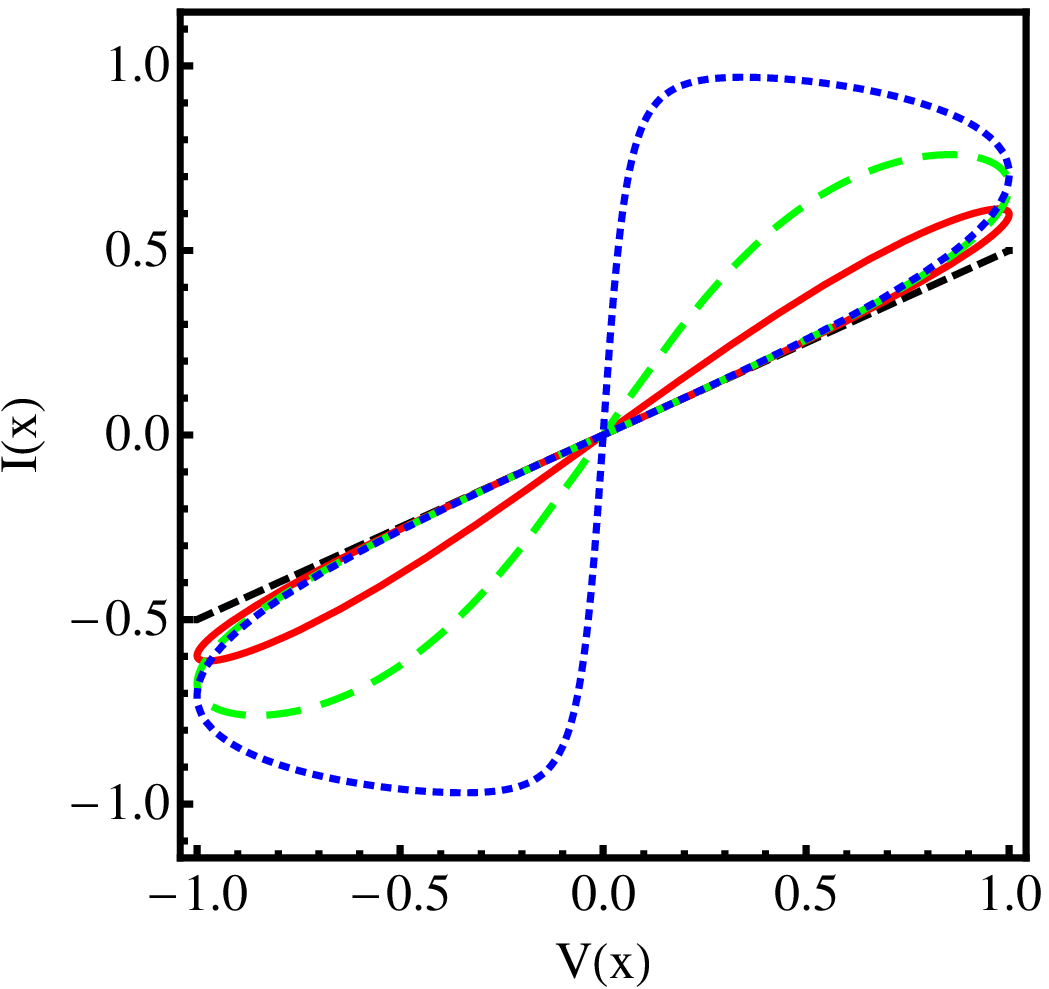}}                
  \subfloat[]{\label{fig_I_V_b}\includegraphics[width=0.33\textwidth]{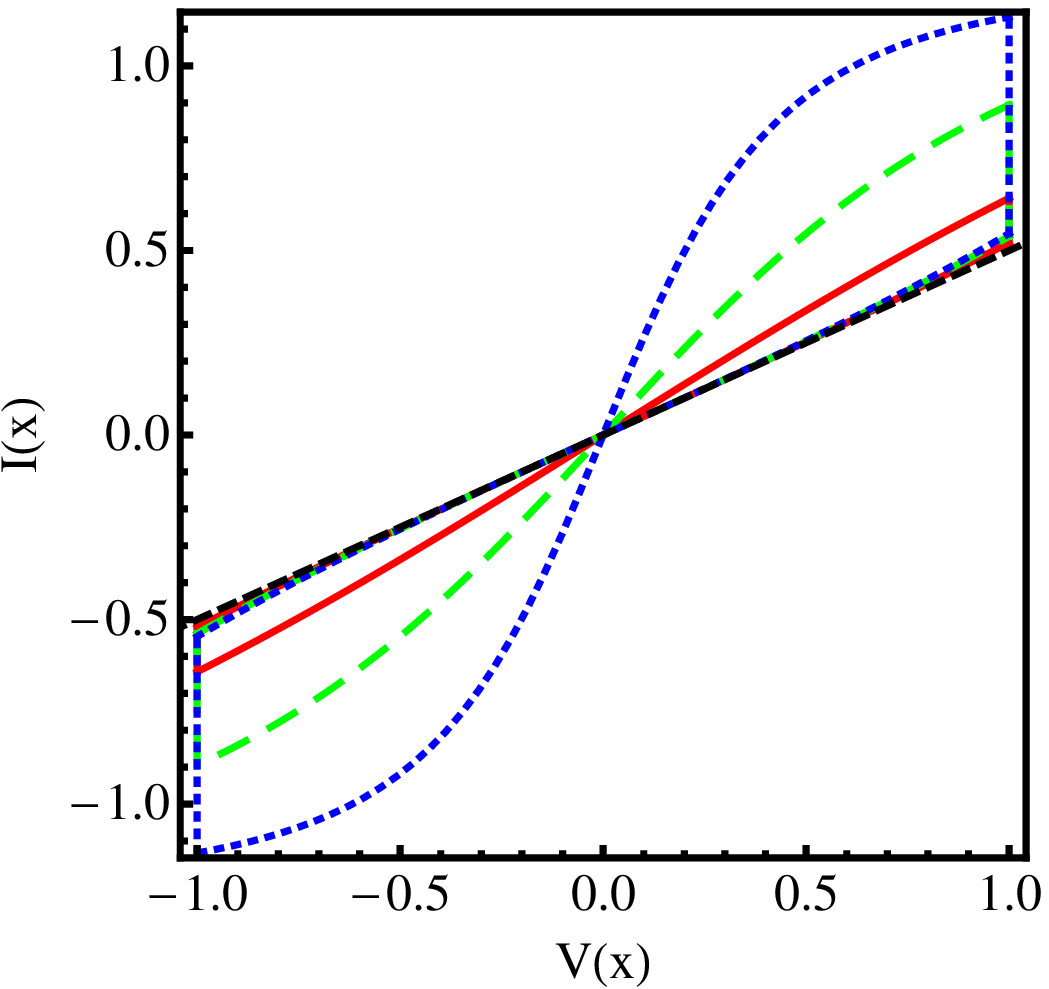}}
  \subfloat[]{\label{fig_I_V_c}\includegraphics[width=0.33\textwidth]{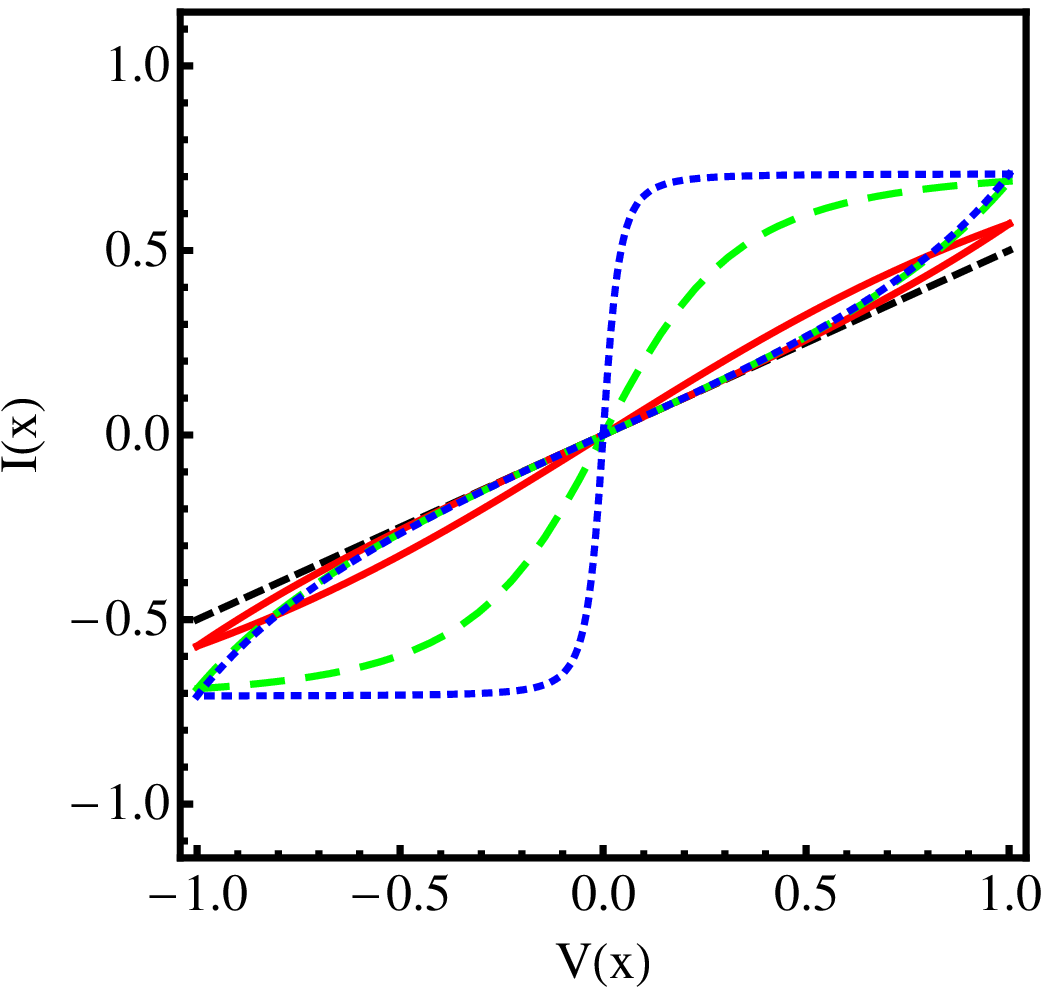}}
  \caption{ Input-output characteristics of the HP memristor under three inputs (arranged by columns) as given in Eqs.~\eqref{eqn_sec2_out_sin_x}--\eqref{eqn_sec2_out_triang_x}.
 {\bf (a)-(c)  } The three input voltages as a function of the normalized time $x$: (a)  sinusoidal input $\sigma(x)$; (b) bipolar piecewise linear wave $\sqcap_{m=1/8}(x)$; and (c) triangular wave $\wedge(x)$.
 \textbf{(d)-(f)} The effect of the parameter $\beta$ on the output current of the HP memristor for the three inputs.  In each figure, the normalized output of the device at different values of $\beta$ (inset) is represented together with the normalized output of an ohmic (linear) resistor (black dashed line). As $\beta$ approaches its upper bound $\beta_{max}$, given in Eq.~(\ref{eqn_sec2_b_boundaries}), the output becomes more nonlinear and deviates from the linear response. 
 \textbf{(g)-(i)} The effect of $\beta$ on the $i-v$ characteristics of the HP memristor for the three inputs. Each figure represents the plot of $[v(x),i(x)]$ for different $\beta$ values together with the linear response of the ohmic resistor (black dashed line). The value of the linear resistor used here is $R_0=2\Omega$. As $\beta \to \beta_{max}$, the $i-v$ becomes more hysteretic. 
 }
 \label{fig_I-V-x_ALL}
 \end{center}
\end{figure*}

We can use the explicit solution~(\ref{eqn_sec2_Williams1_solution}) to calculate the output response of the HP memristor to different voltage drives. Here, we will study periodic input signals $v(t)$ with period $T_0=2\pi / \omega_0$, amplitude $A$, and zero mean value, $\frac{1}{T_0} \int_0^{T_0} v(\tau) d\tau = 0$. Such inputs are standard test signals for electronic devices and lead to the classic characterization of the memristor through hysteresis and Lissajous-type $i-v$ curves~\cite{Chua1971}. In particular, we consider three test signals widely used in engineering settings, namely, the sinusoidal, bipolar piece-wise linear and triangular waveforms, all defined over a period $0<t\leq T_0$ as follows:
\begin{enumerate}
 \renewcommand{\theenumi}{(\roman{enumi})}
 \renewcommand{\labelenumi}{\theenumi}
 \item Sinusoidal voltage input:
  \begin{equation}
  \label{eqn_sec2_sinIn_def}
    \sigma(t) = A\sin(\omega_0t),
  \end{equation}
 \item Bipolar piece-wise linear wave:
  \begin{equation}
  \label{eqn_sec2_sqrIn_def}
  \sqcap_m(t) = 
  \left\{
   \begin{array}{ll}
   \frac{A}{m} \frac{t}{T_0}                           & 0             \leq \frac{t}{T_0} < m \\
   A                                         & m             \leq \frac{t}{T_0} < \frac{1}{2}-m \\
   -\frac{A}{m}(\frac{t}{T_0} - \frac{1}{2}) & \frac{1}{2}-m \leq \frac{t}{T_0} < \frac{1}{2}+m\\
   -A                                        & \frac{1}{2}+m \leq \frac{t}{T_0} < 1 - m\\
   \frac{A}{m}(\frac{t}{T_0} - 1)            & 1-m           \leq \frac{t}{T_0} < 1\\
  \end{array} \right.,
\end{equation}
  \item Triangular wave:
  \begin{equation}
   \label{eqn_sec2_triIn_def}
	\wedge(t) = \sqcap_{m=1/4}(t).
  \end{equation}
\end{enumerate}
For the bipolar piece-wise linear wave, the parameter $0 < m \leq 1/4$ determines the rise and fall time, assumed equal here. In particular, $m=0$ corresponds to the (Heaviside) square wave, while $m=1/4$ is the triangular wave~(\ref{eqn_sec2_triIn_def}). The three input waveforms are shown in Figs.~\ref{fig_Ix_Vx_R0_a}-\ref{fig_Ix_Vx_R0_c}. 

It is convenient to normalize the $i-v$ characteristics~(\ref{eqn_sec2_Williams1_solution}) in terms of rescaled variables:
\begin{equation}
 \label{eq:renormalized}
 \begin{array}{ll}  
\text{a normalized time (or phase):} & x=\omega_0 \, t \\
\text{a normalized input:}  & \widehat{v}=v/A  \\  
\text{a normalized output:} &\widehat{i}=i R_{0}/A,
\end{array}
\end{equation}
to give
\begin{equation}
\label{eq:IVnormalized}
\widehat{i} = \frac{\widehat{v}(x)}{\sqrt{1 - \beta \int_0^{x} \widehat{v}(x)dx}},
\end{equation}
where we have defined $\beta$, a positive dimensionless quantity which combines the parameters of the device and of the input drive:
\begin{equation}
\label{eqn_sec_2_beta_def}
 \beta =
 \frac{2 A}{\omega_0 R_0^2} \, \mu_v \left(\frac{\mathcal{R}_{ON}}{D}\right)^2 \left(\frac{\mathcal{R}_{OFF}}{\mathcal{R}_{ON}}-1\right) = 
 \frac{2 A \, \kappa}{\omega_0 R_0^2}.
\end{equation}
Note that the renormalized parameter $\beta$ includes material and fabrication properties of the device ($\mu_v$, $\mathcal{R}_{ON}$, $\mathcal{R}_{OFF}$, $D$); parameters dependent on the preparation of the state of the device ($R_0$); as well as properties of the driving signal ($A$, $\omega_0$). We will show below that the memristive behavior of the device is fully encapsulated by the parameter $\beta$.

Using the solution~(\ref{eq:IVnormalized}), the response to inputs~(\ref{eqn_sec2_sinIn_def})--(\ref{eqn_sec2_triIn_def}) over a period  $0 \leq x \leq 2 \pi$ is:
\begin{enumerate}
 \renewcommand{\theenumi}{(\roman{enumi})}
  \renewcommand{\labelenumi}{\theenumi}
 \item Sinusoidal voltage input:
\begin{equation}
\label{eqn_sec2_out_sin_x}
\widehat{i}= \frac{\sin(x)}{\sqrt{1 - \beta (1-\cos(x))}}    \quad 0\leq \tfrac{x}{ 2 \pi}  < 1
\end{equation}
\item Bipolar piece-wise linear wave:
\begin{equation}
\label{eqn_sec2_out_square_x}
\widehat{i}=
\renewcommand{\arraystretch}{2} 
\left\{
 \begin{array}{ll}
  \dfrac{\tfrac{1}{m} \tfrac{x}{2\pi}}{\sqrt{1-\tfrac{ \beta \pi}{m}g_1(x)}}     \hspace*{.1in} &0 \leq \frac{x}{2\pi} < m\\
  \dfrac{1}{\sqrt{1- \beta \pi g_2(x)}}                                 &        m\leq \frac{x}{2\pi} < \frac{1}{2}-m \\
  \dfrac{\frac{1}{m}(\frac{1}{2}-\frac{x}{2 \pi})}{\sqrt{1- \frac{\beta\pi}{m}g_3(x)}}      & \frac{1}{2}-m \leq \frac{x}{2\pi} < \frac{1}{2}+m\\
  \dfrac{-1}{\sqrt{1- \beta\pi g_4(x)}}                            & \frac{1}{2}+m \leq \frac{x}{2\pi} < 1-m\\
  \dfrac{\tfrac{1}{m} (\frac{x}{2\pi}-1)}{\sqrt{1-\frac{\beta\pi}{m}g_5(x)}}    & 1-m \leq \frac{x}{2\pi} < 1,\\
 \end{array} \right. 
\end{equation}
\begin{eqnarray*}
g_1(x) &=& (\tfrac{x}{2 \pi})^2 \\
g_2(x) &=& (\tfrac{x}{\pi}-m) \\
g_3(x) &=& (m-2m^2-\tfrac{x^2}{4\pi^2}+\tfrac{x}{2\pi}-\tfrac{1}{4}) \\
g_4(x) &=& (2-m-\tfrac{x}{\pi}) \\
g_5(x) &=& \left(\tfrac{x}{2\pi}-1\right)^2 \\
\end{eqnarray*}
\item Triangular wave:
\begin{equation}
\label{eqn_sec2_out_triang_x}
\widehat{i}=
 \renewcommand{\arraystretch}{2} 
 \left\{
 \begin{array}{ll}
  \dfrac{2\tfrac{x}{\pi}}{\sqrt{1-\beta \frac{x^2}{\pi}}} & 0 \leq \tfrac{x}{2 \pi} < \frac{1}{4} \\
  \dfrac{2 (1-\frac{x}{\pi})}{\sqrt{1-\beta (2x-\frac{x^2}{\pi}-\frac{\pi}{2})}} & \frac{1}{4} \leq  \tfrac{x}{2 \pi} < \frac{3}{4} \\
  \dfrac{2 (\frac{x}{\pi}-2)}{\sqrt{1-\beta(\frac{x^2}{\pi}-4x+4\pi)}} & \frac{3}{4} \leq  \tfrac{x}{2 \pi} < 1 
 \end{array} \right.
\end{equation}
\end{enumerate}

Figure~\ref{fig_I-V-x_ALL} shows the response of the HP memristor for different values of $\beta$ under the three periodic inputs shown in Figures~\ref{fig_Ix_Vx_R0_a}-\ref{fig_Ix_Vx_R0_c}. Figures~\ref{fig_Ix_Vx_a}-\ref{fig_Ix_Vx_c} show the (nonlinear) output current from the HP memristor for three values of $\beta$, with the more positive values having a larger deviation from the linear output from an ohmic (linear) resistor with resistance $R_0$ (dashed black lines). The response dynamics give rise to Lissajous-type $i-v$ characteristics, which are typical of memristors when driven by a periodic excitation with zero mean value~\cite{Chua1971}. In such cases, the output is a double-valued function that forms a simple closed curve with no self intersections except at the origin. Equivalently, this corresponds to a hysteresis loop crossing the origin in the $i-v$ plane~\cite{Chua1976,Chua1980}. As our results show, the amount of hysteresis in the device is dependent on $\beta$, which changes as a combination of different experimental factors. Figures~\ref{fig_I_V_a}-\ref{fig_I_V_c} illustrate the effect of $\beta$ on the $i-v$ characteristics: the hysteretic behavior of the device is enhanced for the most positive values of $\beta$. Again, the dashed black line is the limiting case when the memristor tends to a linear resistor $R_0$, which corresponds to the limiting case of $\beta=0$. 

\subsection{The lumped parameter $\beta$ encapsulates the dynamical response of the HP memristor} 
\label{The renormalized parameter}

Our analysis above shows that the dynamical response of the HP memristor is governed by the lumped parameter $\beta$. This dimensionless parameter represents in a single quantity the collective effect of device parameters of diverse origin including: material properties, such as carrier mobility and doping ratio ($\mathcal{R}_{ON}$, $\mathcal{R}_{OFF}$, $\mu_v$); device fabrication, such as the depth of the thin-film layer ($D$); preparation of the initial state of the device ($R_0$); as well as properties of the driving input ($A$, $\omega_0$). Therefore, one does not need to consider each individual parameter separately---our analytical expressions show that similar responses can be achieved by changing different properties of the device if they lead to the same value of $\beta$. Furthermore, for a given fabricated device one can tune the properties of the input drive in the experiment to produce particular output responses.

We now examine some of the characteristics of the parameter $\beta$. >From the definition~(\ref{eqn_sec_2_beta_def}), it follows that $\beta > 0$  since $\mathcal{R}_{OFF} > \mathcal{R}_{ON}$. As $\beta \to 0$, the memristor tends to the behavior of a linear resistor with history-dependent resistance $R_0$. Hence, for a given fabricated device with particular physical characteristics, the linear behavior will be revealed experimentally in the limit of low amplitude/high frequency drive ($A \to 0$ and/or $\omega_0 \to \infty$). In the process of designing memristors, linear behavior will be more likely when the dimension of the device is large ($D \to \infty$); when the mobility of the carriers is low ($\mu_v \to 0$); when the doping ratio is small  ($\mathcal{R}_{ON}/\mathcal{R}_{OFF} \to 1$), or with a combination of all of those. In fact, note that $\beta \equiv 0$ when $\mathcal{R}_{ON} = \mathcal{R}_{OFF}$. This corresponds to building an ohmic resistor, rather than a memristor, with just one (undoped) region. In this case, the resistor always responds linearly with the same unique resistance {\it independently} of history or preparation protocol, i.e., $R_0 =  \mathcal{R}_{OFF}$.

In the case of the HP model, the value of $\beta$ is also bounded from above, as follows from requiring that the arguments in the square roots in Eqs.~(\ref{eqn_sec2_out_sin_x})--(\ref{eqn_sec2_out_triang_x}) be positive, so that the output current is real and finite. This leads to the following bounds for $\beta$ for each type of input:
\begin{equation}
 \label{eqn_sec2_b_boundaries}
 \left\{
 \begin{array}{ll}
    0 < \beta < 1/2  & \text{for }  v(t)=\sigma(t)  \\
    0 < \beta < 1/(\pi- 2 \pi m) & \text{for } v(t)=\sqcap(t) \\
    0 < \beta < 2/\pi & \text{for   }v(t) =\wedge(t).  
 \end{array}\right.
\end{equation}
In order to make the comparison across input drives more direct, it is thus helpful to define the following rescaled parameter:
\begin{equation}
 \label{eqn_sec2_g_mapping}
 \widetilde{\beta} \equiv \left\{
 \begin{array}{ll}
    2 \beta & \text{for } v(t)=\sigma(t) \\
    (\pi - 2\pi m) \beta & \text{for }v(t)=\sqcap(t) \\
    (\pi/2) \beta & \text{for }v(t)=\wedge(t),
 \end{array}\right.
\end{equation}
such that $\widetilde{\beta} \in (0,1)$ for all three drives. When $\widetilde{\beta}=0$, the memristor becomes a linear resistor, while as $\widetilde{\beta} \to 1$, the HP memristor becomes maximally nonlinear and hysteretic and the separation of the two branches in the $i-v$ plane is maximized.  In the next section, we make this notion more precise through the introduction of a quantitative measure of hysteresis.

It should be noted that the upper bound of $\beta$, that stems from the requirement that the output remains real and finite, is a manifestation of a limitation of the first simplified HP memristor model~\cite{Strukov2008}, i.e., that   the device does not saturate or break down as the memristance is driven towards its lower bound $\mathcal{R}_{ON}$. Although such a mechanism is not accounted for by Eqns.~(\ref{eqn_sec2_williams1_v}) and (\ref{eqn_sec2_Williams1_w}), this limitation has already been addressed by models that contain a windowing function, in which it becomes increasingly difficult to change the memristance as the boundary between the doped and undoped region reaches either of the two ends of the device~\cite{Strukov2008,Yogesh2009,Pershin2011}.

\section{Quantitative measure of hysteresis}

\subsection{Definition}

Controlling and designing the hysteretic pinch of the $i-v$ curve of the memristor is crucial for the use of these devices both individually or as part of larger circuits. It is important, therefore, to have a method to control the hysteretic effect of a particular model in order to achieve the required specifications for an application. We propose here a quantitative measure of the hysteresis of the $i-v$ curve in terms of the work carried out by the driving input on the device. We will then use the analytical solutions obtained for the HP model in the previous section to calculate the hysteresis of the model for the three different drives in terms of the parameter $\widetilde{\beta}$.  Finally, we will show how these expressions may be used as an aid to fabricate a memristor with a prescribed $i-v$ curve.

Our measure of hysteresis is based on the calculation of the work done by the input signal through the device. Let $H$ denote the positive scalar quantity measuring the difference between the work done while traversing the upper and lower branches of the hysteresis loop in the $i-v$ plane:
\begin{align}
&H = W_{+} - W_{-} = \int_{t_1}^{t_2} i\, v \, dt - \int_{t_2}^{t_3} i\, v\, dt \nonumber \\
&=\frac{A^2}{R_0 \omega_0} \left( \int_{x_1}^{x_2}\widehat{i}\, \widehat{v} \, dx - \int_{x_2}^{x_3}\widehat{i} \, \widehat{v} \, dx \right ) \equiv \frac{A^2}{R_0 \omega_0} \widehat{H},
 \label{eqn_sec3_hysteresis_def}
\end{align}
where $t_1$ to $t_2$ and $t_2$ to $t_3$ is the time required to move along the upper and lower branch of the hysteresis loop, respectively.    Here, $P(t)=i(t)v(t)$ is the instantaneous power and, as it is customary by now, $\widehat{H}$ corresponds to the rescaled quantity. 

Clearly, $H$ is defined in terms of the energy dissipated by the device and becomes zero when the memristor tends to a linear resistor since $W_{+}$ and $W_{-}$ coincide in that case. Therefore, it is useful to scale the hysteresis by $W_0$, the work done on the linear resistor $R_0$ when driven for one complete cycle by the same signal for which $H$ is evaluated:
 \begin{align}
 \label{eqn_sec3_normWork_def}
 W_0 &= \frac{1}{R_0}\int_{T_0} v^2(\tau) d\tau = R_0 \int_{T_0} i^2(\tau) d\tau \\
     &= \frac{A^2}{R_0 \omega_0} \int_{2\pi} \widehat{v}^2 \,dx \equiv \frac{A^2}{R_0 \omega_0} \widehat{W}_0.
\end{align}
We define the scaled hysteresis as:
\begin{equation}
 \label{eqn_sec3_normHysteresis_def}
 \bar{H} \equiv \frac{H}{W_0} =  \frac{\widehat{H}}{\widehat{W}_0}.
\end{equation}

\subsection{Analytical expressions for the hysteresis of the HP memristor}

We now use our analytical results for the HP model to obtain the hysteresis of this device under the three drives specified in the previous section. 
\subsubsection{Sinusoidal drive:}
Consider an HP memristor with sinusoidal input $v(t)=\sigma(t)$ and the corresponding output current $\widehat{i}$,  given by Eq.~(\ref{eqn_sec2_out_sin_x}). The hysteresis $\widehat{H}$ is then:
\begin{eqnarray}
 \label{eqn_sec3_hysteresis_sin}
 \widehat{H} &=& 2 \left[ \int_{\frac{\pi}{2}}^{\pi} \frac{\sin^2(x)}{\sqrt{1-\beta (1- \cos(x))}} dx \right. \\ 
   & & \qquad \qquad  \qquad  \left. - \int_{0}^{\frac{\pi}{2}} \frac{\sin^2(x)}{\sqrt{1-\beta (1 - \cos(x))}}dx \right]. \nonumber
   \label{eqn:evalH:sinusoid}
\end{eqnarray}
The work done by the input  signal $\sigma(t)$ for a complete cycle on the linear resistor $R_0$ is:
\begin{equation*}
 \label{eqn_sec3_normWork_sin}
 \widehat{W}_0 = \int_0^{2\pi} \sin^2(x) dx = \pi.
\end{equation*}
Using the symmetry of the functions, the scaled hysteresis $\bar{H}$ is given by:
\begin{multline}
 \label{eqn_sec3_normHysteresis_sin_step1}
  \bar{H} = \frac{2}{\pi} \left[ \int_{0}^{\pi} \frac{\sin^2(x)}{\sqrt{1-\beta (1-\cos(x))}} dx \right. - \\ 
            \left. 2\int_{0}^{\frac{\pi}{2}} \frac{\sin^2(x)}{\sqrt{1-\beta (1-\cos(x))}}dx \right],
\end{multline}
which can be rewritten as~\cite{Elliptic_integrals,Table_of_integrals}: 
\begin{align}
 \label{eqn_sec3_normHysteresis_sin_step3}
 \bar{H} & = \frac{32\sqrt{1 -\widetilde{\beta}}}{3\pi\widetilde{\beta}^2} \Bigg\{ \left[2F(\phi_2,k) - K(k)\right]  \nonumber\\
 &        + \left(1-\frac{\widetilde{\beta}}{2}\right) \left[E(k) - 2E(\phi_2,k)\right] -\frac{\widetilde{\beta}}{2}\sqrt{\frac{1-\tfrac{\widetilde{\beta}}{2}}{1-\widetilde{\beta}}} \Bigg\}   
\end{align}
where $k^2=\frac{\widetilde{\beta}}{\widetilde{\beta}-1}$ and  $\phi_2=\arcsin\sqrt{\frac{1-\widetilde{\beta}}{2-\widetilde{\beta}}}$. Here, $F(\phi,k)$ and $E(\phi,k)$ are the incomplete elliptic integrals of the first and second kind, respectively, while $K(k)$ and $E(k)$ denote the complete elliptic integrals of the first and second kind, respectively~\cite{Elliptic_integrals,Table_of_integrals}. 
\begin{figure}[!hb]
 \begin{center}
    \includegraphics[width=0.4\textwidth]{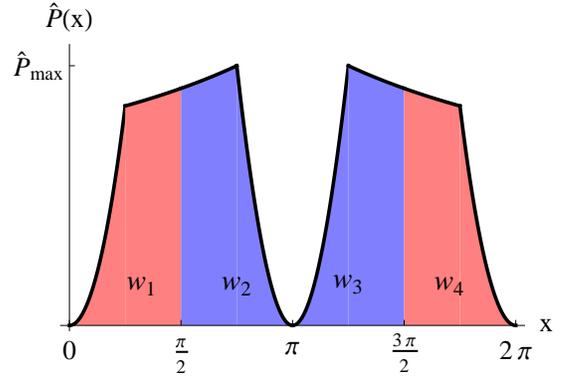}
  \caption{The instantaneous power, $\widehat{P}=\widehat{i} \,\, \widehat{v}$, as a function of the normalized time $x$ for  the HP memristor when driven by the bipolar square wave $\sqcap_{m=1/8}(x)$. The output current is given by Eq.~(\ref{eqn_sec2_out_square_x}). The areas $w_1$ to $w_4$ indicate the work for the time intervals  $[0,\pi/2)$, $[\pi/2,\pi)$, $[\pi,3\pi/2)$ and $[3\pi/2,2\pi]$, respectively. The hysteresis~(\ref{eqn_sec3_hysteresis_def}) is equivalent to $H=W_+ - W_-= (w_2+w_3) - (w_1 + w_4)$ where $W_+$ and $W_-$ indicate the work along the upper and lower branches, respectively.  The asymmetry in the instantaneous power is a reflection (and a measure) of the observed hysteresis.}
  \label{fig_sqr_Power}
\end{center}
\end{figure}

\begin{figure*}[!ht]
 \begin{center}
  \subfloat[]{\label{fig_Hnorm-g_ALL}\includegraphics[width=0.5\textwidth]{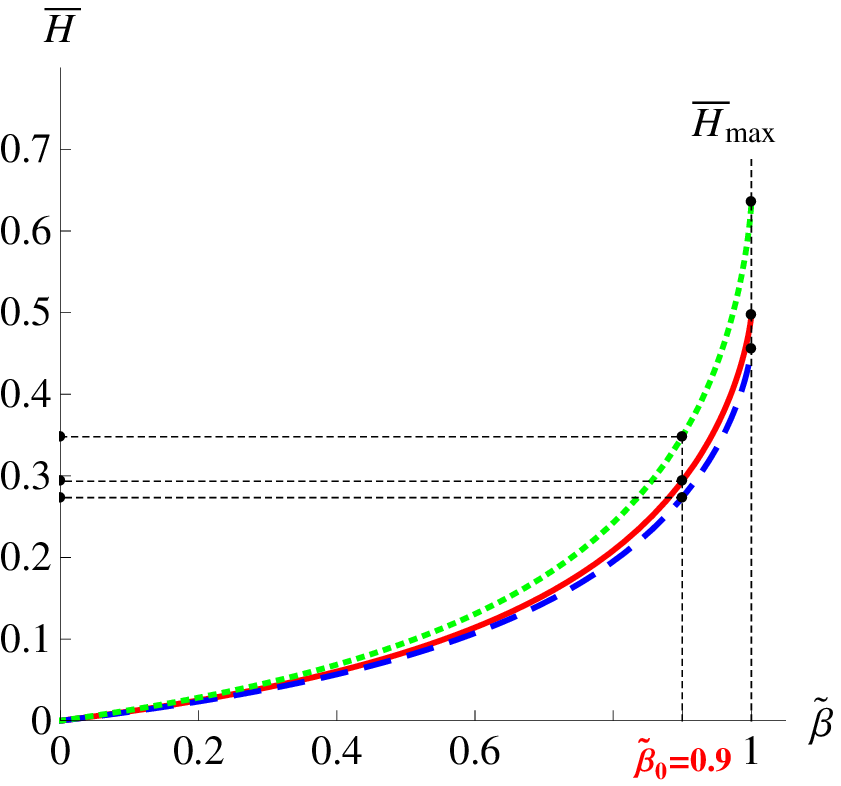}} \hfill         
  \subfloat[]{\label{fig_I_V_ALL}\includegraphics[width=0.45\textwidth]{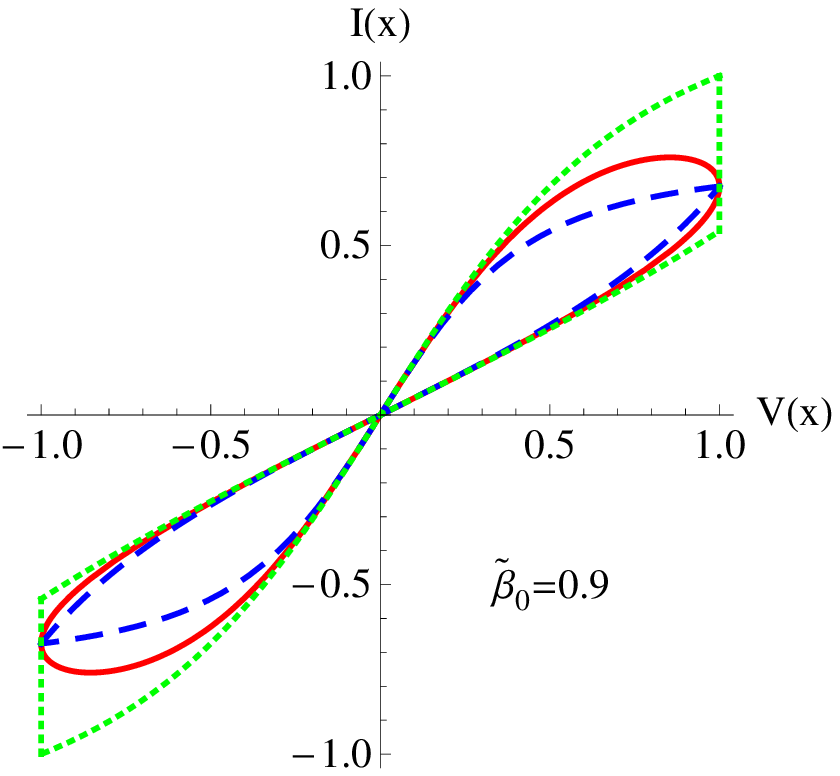}}
  \caption{(a) The normalized hysteresis $\bar{H}$ of the HP memristor model as a function of the memristor lumped parameter $\widetilde{\beta}$ under: a sinusoidal input (red solid line, Eq.~(\ref{eqn_sec3_normHysteresis_sin_step3})), a bipolar piecewise linear drive with $m=1/20$ (green dotted line, Eq.~(\ref{eqn_sec3_normHysteresis_sqr})) and a triangular input (blue dashed line, Eq.~(\ref{eqn_sec3_normHysteresis_triang})). The hysteresis curves achieve finite maxima $\bar{H}_{\mathrm{max}}$ at $\widetilde{\beta}=1$. For a given lumped parameter (e.g., $\widetilde{\beta}_0 =0.9$), we can obtain the hysteresis of the memristor when driven by any of the three inputs studied in Fig.~\ref{fig_I-V-x_ALL}. The corresponding $i-v$ characteristics of the HP model with $\widetilde{\beta}_0 =0.9$ are shown in (b).  Conversely, one can evaluate the hysteresis  from $i-v$ curves, obtained either theoretically or experimentally, and determine the corresponding $\widetilde{\beta}$.}
  \label{test}
 \end{center}
\end{figure*}
\subsubsection{Bipolar piecewise linear input}
After some algebra, the scaled hysteresis of the bipolar input waveform $v(t)=\sqcap(t)$ is obtained as~\cite{Table_of_integrals}:
\begin{multline}
 \label{eqn_sec3_normHysteresis_sqr}
 \bar{H} = \frac{3\sqrt{m'}}{\widetilde{\beta}(3-8m)}\Biggl\{\frac{1}{\sqrt{\widetilde{\beta} m}}\Biggr[ -m' \arcsin\sqrt{\frac{\widetilde{\beta} m }{m'}} \\   
  + m' \ln \left( \frac{\sqrt{m'+\widetilde{\beta}(m-m')}-\sqrt{\widetilde{\beta} m}}{\sqrt{m'(1-\widetilde{\beta})}}\right)^{1-\widetilde{\beta}}\\
  +\sqrt{\widetilde{\beta} m}\left[\sqrt{m'+\widetilde{\beta}(m-m')}+\sqrt{m'-\widetilde{\beta} m} \right] \Biggl] \\
  + 2\Biggl( 2\sqrt{m'(1- \frac{\widetilde{\beta}}{2}}) -\sqrt{m'-\widetilde{\beta} m} \\
  - \sqrt{m' + \widetilde{\beta}(m-m')} \Biggr) \Biggr\},
\end{multline}
where $m'=1-2m$.
\subsubsection{Triangular wave input}
The hysteresis for the triangular wave input $v(t)=\wedge(t)$ is obtained by particularizing Eq.~(\ref{eqn_sec3_normHysteresis_sqr}) for $m=1/4$:
\begin{multline}
 \label{eqn_sec3_normHysteresis_triang}
 \bar{H} = \frac{3}{\sqrt{2 \, \widetilde{\beta}^{3}}} \Biggl[ \left(1-\widetilde{\beta} \right) \ln\left( \frac{\sqrt{2-\widetilde{\beta}}-\sqrt{\widetilde{\beta}}}{\sqrt{2(1-\widetilde{\beta})}} \right)\\
 -\arcsin\sqrt{\frac{\widetilde{\beta}}{2}}+\sqrt{\widetilde{\beta}(2-\widetilde{\beta})}\Biggr].
\end{multline}

\subsection{The dependence of the hysteresis on the parameter $\beta$}
The definition of hysteresis~(\ref{eqn_sec3_hysteresis_def}) is based on the integration of the instantaneous power consumed by the device over the course of an input cycle. Figure~\ref{fig_sqr_Power} shows a plot of the instantaneous power for the HP model driven by the bipolar wave $\sqcap(t)$. Each of the areas $w_1$ to $w_4$ indicate the work done by the input signal for the corresponding time interval.  The hysteresis is given by $H=(w_2+w_3) - (w_1 + w_4)$ where $W_+=w_2+w_3$ and $W_-=w_1+w_4$.  Note that $w_2=w_3\neq w_1=w_4$. This asymmetry is a reflection of the difference in the work carried out on each of the two branches of the $i-v$ curve due to the nonlinearity of the device and giving rise to the hysteresis.
 
The expressions of the memristor hysteresis subject to the three drives given  by Eqs.~(\ref{eqn_sec3_normHysteresis_sin_step3})-(\ref{eqn_sec3_normHysteresis_triang}) are all explicit functions of $\widetilde\beta$, the lumped parameter that combines the physical parameters of the device together with the properties of the drive. Figure~\ref{fig_Hnorm-g_ALL} shows that for all drives, the hysteresis of the device is zero when $\widetilde \beta = 0$ and  increases as $\widetilde\beta \to 1$. It is important to remark that the device has a \textit{finite} maximum value of hysteresis it can exhibit, i.e., the value of $\bar{H}$ does {\it not} diverge as $\widetilde \beta \to 1$.  In fact, we can use our analytical expressions to calculate the upper bound of the hysteresis for each drive. For instance, from Eq.~(\ref{eqn_sec3_normHysteresis_sin_step1}) the maximum hysteresis for the HP memristor driven by a sinusoidal input is:
\begin{align*}
\bar H_{\mathrm{max}}&\equiv\bar H({\widetilde\beta=1})= \frac{2}{\pi} \left[ \int_{\frac{\pi}{2}}^{\pi} \frac{\sin^2 x}{\cos(\tfrac{x}{2})} dx  - 
            \int_{0}^{\frac{\pi}{2}} \frac{\sin^2 x}{\cos(\tfrac{x}{2})}dx \right] \nonumber \\
            &= \frac{16}{3 \pi} \left(1-\frac{1}{\sqrt{2}}\right)
 \approx 0.4972,
\end{align*}
while for the triangular input, the upper bound of the hysteresis is slightly lower:
\begin{equation*}
\bar{H}(\widetilde\beta=1)= \frac{3}{\sqrt{2}} \left(1-\frac{\pi}{4} \right) \approx 0.4552.
\end{equation*}
Therefore, the maximum hysteresis exhibited by the HP memristor, understood as the difference between the work along the upper and lower branches of the $i-v$ over a period of the input, is equivalent to $\sim 50\%$ of the energy dissipated by the equivalent ohmic resistor.

The dependence of the hysteresis on the lumped parameter $\beta$ can be used during the design process of a memristor or during the characterization of a fabricated device. If the aim is to design a memristor with a pre-specified $i-v$ response that needs to operate under a particular input, one may calculate first the hysteresis $H$ of the desired $i-v$ curve and the value of $\beta$ that will produce the desired response. The identified $\beta$ value can then be used to restrict our fabrication parameters. Similarly, for a given fabricated memristor, one can generate different $i-v$ characteristics under a particular type of drive with varying frequency and/or amplitude, and/or under a different type of drive. For each $i-v$ curve, the scaled hysteresis $\bar{H}$ can be obtained from the experimental data. If the data is well described by the HP model, our expressions could be used to fit some of the intrinsic parameters of the device that are contained in $\beta$. The interplay between hysteresis and $\beta$ is exemplified in Fig.~\ref{fig_I_V_ALL} for the $i-v$ curves generated by a memristor with $\beta_0=0.9$ under the three input drives.

\section{Discussion}

In this paper, we have presented a mathematical framework for the analysis of the input-output dynamics of memristors. Although these are nonlinear elements, the form of the constitutive relation of the memristor leads in general to Bernoulli dynamics,  which can be directly linked to an associated linear differential equation through a nonlinear transformation. Table~\ref{tbl_ch3_general_solutions} highlights this general form of memristor dynamics and the duality that emerges when devices with distinct internal control variables are interrogated with different externally controlled variables, i.e., charge- and flux-controlled memristances which can be voltage- and current-driven.

Our methodology allows us to obtain,  in some cases, the output of the device as an explicit function of the input. We exemplified our analysis with the recently introduced HP memristor model~\cite{Strukov2008}, which is shown to be a Bernoulli memristor, for which we obtain \textit{analytically} the output current explicitly as a function of the voltage for three typical input signals, namely, the sinusoidal, bipolar piecewise linear and triangular waveforms.  Our analysis led us to the identification of the dimensionless lumped parameter $\beta$, which combines physical parameters of the device as well as properties of the input drive, and controls the hysteretic properties of the $i-v$ characteristics.  We have shown elsewhere~\cite{Georgiou2011} that the same parameter also quantifies the amount of nonlinearity in the output spectrum of the device. Consequently, $\beta$ controls both the hysteresis of the memristor and the harmonic distortion that the device introduces. The functional form~\eqref{eq:IVnormalized} makes apparent how $\beta$ controls the memristive character of the device and reveals the fundamental interlinking between nonlinearity, hysteresis and memory in these devices.

The explicit solutions thus obtained can be of use not only for the understanding of memristive dynamics but also as a means to parameterize experiments and to study the deviation from idealized models. They can also be used to devise experiments that can reveal the memristive properties of particular devices. For instance, it is possible to use $\beta$ to design an input drive (e.g., its functional form, amplitude and frequency) that will enhance the hysteretic response of a particular memristor, dependent on the physical properties of the device and the underlying transport mechanism of the charge carriers.  The parameter $\beta$ also indicates which fabrication parameters are likely to enhance (or reduce) memristive behavior. In the HP model, the small dimensions of the device, a large carrier mobility, and large differences between the doped and undoped resistances all contribute to make the memristive behavior more conspicuous.  The results presented could be applicable to experiments given that the simple HP model studied here has already been shown to provide a reasonable approximation of the behavior of particular experimental realizations~\cite{Pinaki2010}.

Although we have focused here on particular examples, this analytical approach has been extended to other memristor models~\cite{Cai2011}. Furthermore, equivalent analytical input-output relations can be obtained for {\it networks} of memristors and other mem-elements connected to circuit parasitics~\cite{Georgiou2011}.  The study of such directions as well as the investigation of more realistic memristor models (in particular those that account  for non-linear dopant kinetics) will be the object of further research.

\bibliography{PR-B}

\end{document}